\documentclass{cup-hpl}

\begin{document}
\bibliographystyle{unsrtnat}




\title{A comprehensive study of second and third harmonic conversion efficiency, angular and temperature tolerance, and long-term stability in LBO crystals using a 10-J-class laser}

\author[1,2]{Huzefa Aliasger}
\author[1]  {Ondřej Novák\corresp{Huzefa Aliasger, HiLASE Centre, Institute of Physics of the Czech Academy of Sciences, Za Radnicí 828, 25241 Dolní Břežany, Czechia.   \email{huzefa.bhanpurwala@hilase.cz}}}
\author[1]{Zbyněk Hubka}
\author[1]{Martin Hanuš}
\author[1]{Petr Navrátil}
\author[1]{Patricie Severová}
\author[1,2]{Ondřej Denk}
\author[1]{Jan Pilař}
\author[1,2]{Tomáš Paliesek}
\author[1]{Martin Divoký}
\author[3]{Ondřej Schreiber}
\author[2]{Michal Jelínek}
\author[1]{Martin Smrž}
\author[1]{Tomáš Mocek}

\address[1]{HiLASE Centre, Institute of Physics of the Czech Academy of Sciences, Za Radnicí 828, 25241 Dolní Břežany, Czechia}

\address[2]{
	Department of Laser Physics and Photonics, Faculty of Nuclear Sciences and Physical Engineering, Czech Technical University in Prague, 
	Břehová 7, 115 19 Praha 1, Czechia
}

\address[3]{Narran s.r.o., Bayerova 802/33, Brno 602 00, Czechia}

\begin{abstract}
We present a study of second harmonic generation (SHG) and third harmonic generation (THG) in lithium triborate (LBO) crystals using a high-energy, 10-J-class, 10 Hz Yb:YAG laser system. We achieved high conversion efficiencies of 75\% for SHG and 56\% for THG for Gaussian-like temporal pulse shapes and top-hat-like beam profiles. The angular and temperature dependence of the LBO crystals were measured and validated through numerical simulations. The SHG process exhibited an angular acceptance bandwidth of 1.33 mrad and a temperature acceptance bandwidth of 2.61 K, while the THG process showed 1.19 mrad and 1.35 K, respectively. Additionally, long term stability measurements revealed RMS energy stabilities of 1.3\% for SHG and 1.24\% for THG. These results showcase the reliability of LBO crystals for high-energy, high-average-power harmonic generation. The developed system offers automated switching between harmonics provided at the system output. The system can be easily adapted to Nd:YAG based pump lasers as well.
\end{abstract}

\keywords{frequency conversion, high-energy, high-average-power, LBO, long term stability, angular acceptance bandwidth, temperature acceptance bandwidth}

\maketitle

\section{Introduction}

The interaction of matter at high-energy, high-power lasers has paved a way to explore opportunities in real world applications such as laser particle acceleration \cite{PhysRevLett.43.267}, medical treatments \cite{PhysRevAccelBeams.27.073501}, precision micromachining \cite{Klotzbach2011}, laser shock peening \cite{KAUFMAN2025108982, Stránský31122024} and laser driven fusion \cite{10.1063/1.2748389, DINICOLA2024101130, PhysRevLett.132.065102}. Ytterbium doped nanosecond (ns) solid-state lasers are of significant interest due to their suitability for high-energy, high-power and high repetition rate applications \cite{Paschottaytterbium_doped_laser_gain_media}. This suitability arises from their millisecond scale fluorescence lifetime, superior thermal conductivity (attributed to their YAG host), and compatibility with diode laser pumping \cite{902180, Divoky:21}. The high-energy pulses from Yb:YAG, which are around 1030 nm, can be used as a pre-amplifier in a Chirped Pulse Amplification (CPA) system \cite{https://doi.org/10.1002/lpor.201400040}. However, they are not suitable for pumping Optical Parametric Chirped Pulse Amplification (OPCPA) systems or Ti:sapphire lasers, which generate femtosecond pulses at 800 nm and require a high-energy pump source around 500 nm \cite{Lyachev:11, Papadopoulos_Zou, Bayramian:08}. Therefore, frequency conversion techniques such as SHG must be used to address this wavelength mismatch, thereby enhancing the utility of high-energy and high-power lasers. Furthermore, high-energy pulses at even shorter ultraviolet (UV) wavelengths are essential for laser annealing of semiconductors \cite{mi15010103}. These UV pulses can be generated through THG of a 1 $\upmu$m laser.

In the past, several groups around the world have succeeded to convert frequency of high-energy pulses at around 1 $\upmu$m, successfully generating high-energy pulses in the green and UV spectra. Phillips et al. \cite{Phillips:16} reported SHG conversion efficiency of 82\% using LBO crystal, achieving 5.6 \nolinebreak J at a repetition rate of 10 Hz from 1030 nm. Also, Philips et al. \cite{Phillips:21} demonstrated SHG with a conversion efficiency of 66\% achieving 59.7 J at 515 nm with a repetition rate of 10 Hz and THG at conversion efficiency of 68\% achieving 65 J at 343 nm at a repetition rate of 1 Hz. Sekine et al. \cite{Sekine:13} achieved 12.5 J at 527 nm using CLBO crystal, with a conversion efficiency of 71.5\% at a repetition rate of 0.6 \nolinebreak Hz. Chi et al. \cite{Chi:20} reported 0.94 J laser pulses at 515 nm with 2 ns duration and a 1 kHz repetition rate, achieving 78\% efficiency using LBO crystal. Divoky et al. \cite{Divoky_Phillips} reported SHG achieving 95 J at 515 nm at 10 Hz, marking the highest energy ever achieved in the green spectrum at this repetition rate. Pilar et al. \cite{Pilar_Divoky} reported THG achieving 50 J at 343 nm with a repetition rate of 10 Hz, marking the highest energy ever achieved at this repetition rate.

In this paper, we present the results of SHG and THG in LBO crystals using the output beam from 10 J Bivoj amplifier operating at a repetition rate of 10 Hz. Unlike the top-hat temporal profile employed by Phillips et al. \cite{Phillips:16}, we utilized Gaussian-like pulses while maintaining a spatially top-hat beam profile, achieving high conversion efficiencies of 75\% for SHG and 56\% for THG. We conducted detailed measurements of the angular and temperature dependencies of the LBO crystals, which were further corroborated by numerical simulations. Furthermore, we characterized the long-term energy stability of the SHG as well as THG process in the LBO crystals.

\section{Experimental setup}

\begin{figure*}[t!]
	\centering
	\includegraphics[scale=0.5]{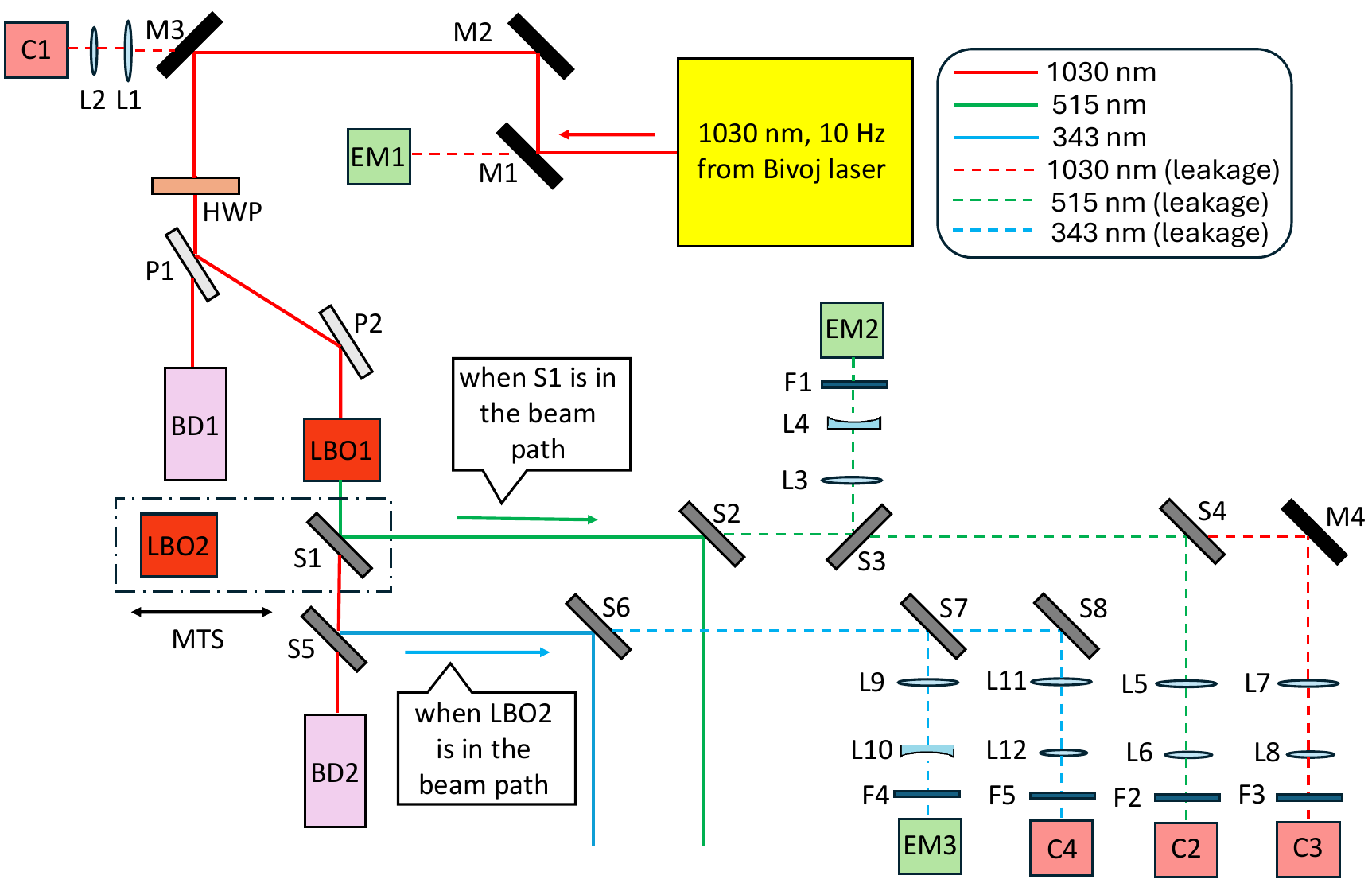}
	\caption{Experimental setup. M: mirrors; HWP: half waveplate; P: polarizers; S: harmonic separators (dichroic mirrors); L: lenses; C: cameras; EM: energy meters; BD: beam dumps; F: filters; MTS: motorized translation stage}
	\label{fig:Experimental_setup}
\end{figure*}

The schematic of the experimental setup is shown in Fig.~\ref{fig:Experimental_setup}. The driving laser beam stems from the 10 J amplifier of the Bivoj laser system \cite{Mason:17, Divoky:21}. The Bivoj laser is a cryogenically cooled Yb:YAG multipass amplifier system that delivers a beam at 1030 nm and operates at a repetition rate of 10 \nolinebreak Hz. The temporal pulse profile was adjusted to provide 8 ns Gaussian-like pulses. We decided to use Gaussian-like pulses for this study because they represent the typical pulse shape of ns lasers. Achieving a top-hat temporal profile requires a specialized front-end configuration, which is not commonly available in standard commercial laser systems and often involves added complexity and cost. The square-shaped, top-hat-like beam from the Bivoj amplifier is relayed to the optical table in the different part of the laboratory with the help of the laser beam distribution system (LBDS) which contains several imaging telescopes.

On the optical table, the laser beam is first directed towards the harmonic generation system using mirrors M1, M2 and M3. These mirrors enable alignment of the input beam. The leakage of the laser beam through mirror M1 is utilized to monitor the input fundamental energy using energy meter EM1 (Ophir PE50-C), while the leakage from mirror M3 is used to monitor the near-field profile of the input fundamental beam via camera C1 after its demagnification by a telescope. The attenuator at the input of the system consists of a half-wave plate HWP mounted on a motorized rotation stage and two thin-film Brewster polarizers P1 and P2. The attenuator is used to control the energy of the laser pulses entering the nonlinear crystals without affecting its spatial or temporal shape. The s-polarized part of the beam is reflected by the polarizers and guided into the nonlinear crystals for harmonic generation. The p-polarized beam is transmitted through P1 polarizer and is directed towards beam dump BD1. 

The second harmonic (SH) beam generated from crystal
LBO1 is separated from the unconverted fundamental beam using harmonic separator S1. The separator is highly reflective for the SH beam and highly transmissive for the fundamental beam. The unconverted fundamental beam, separated from the SH by separator S1 is directed towards beam dump BD2. After separation, the SH is directed to the output via harmonic separators S2 and the leakage from S2 is utilized for diagnostics of both the SH and the unconverted fundamental beam. The leakage through S2 is directed to energy meter EM2 (Ophir PE10-C) for measuring the energy of the SH pulses using reflection from separator S3. The beam size is reduced by a telescope to fit to EM2 aperture.  Additionally, the leakage through S3 is directed onto camera C2 through a demagnifying Keplerian telescope to capture the near-field profile of the SH beam reflected by separator S4. Finally, the leakage through S4 is directed via mirror M4 to camera C3 through a demagnifying telescope to capture the near-field profile of the unconverted fundamental beam. The demagnifying telescopes image the crystal position onto the cameras. Appropriate spectral filters are used in front of the energy meters and cameras to transmit only the harmonic to be observed.

Harmonic separator S1 and LBO2 crystal for THG are mounted on a common motorized translation stage. This setup enables automated switching between the SH and third harmonic (TH) beams, making it well-suited for practical applications, e.g when a wavelength dependence of the interaction is to be studied. When only SH beam is required at the output, S1 is positioned into the beam path. Conversely, LBO2 is placed in the beam path when TH is required. The TH beam generated in LBO2 crystal is separated from the SH and the remaining unconverted fundamental beam using harmonic separator S5, which is highly reflective for the TH beam and highly transmissive for the fundamental and SH beam. The unconverted fundamental and SH beams separated from the TH beam are directed towards beam dump BD2. After separation, the TH is directed to the output by reflection off S6. The leakage through S6 is reflected by S7 and directed to energy meter EM3 (Gentec QE12LP-H-MB-QED-D0) for measuring the energy of the TH pulses. The beam size is reduced by a telescope to match EM3 aperture. Additionally, the beam reflected by S8  is directed through a beam size reducing Keplerian telescope to camera C3 to capture the near-field profile of the TH beam. Appropriate band pass filters are placed in front of energy meter EM3 and camera C4 to filter out unwanted frequencies. The energy meters were calibrated in a way that the additional energy meters were placed into the output beams except EM1 measuring fundamental beam where the energy meter was placed in front of the HWP.

LBO1 crystal used for SHG has an aperture of 30 mm $\times$ 30 mm and a length of 20 mm. The crystal is cut at angles $\theta$ = 90° and $\phi$ = 11.6° for type I (o + o $\rightarrow$ e) SHG in the XY plane. Both faces of the crystal are coated with dual-band anti-reflection (DBAR) coatings for 1064 nm and 532 nm. Similarly, LBO2 crystal used for THG also has an aperture of 30 mm $\times$ 30 mm and a length of 20 mm. This crystal is cut at angles $\theta$ = 42.4°and $\phi$ = 90° for type II (o + e $\rightarrow$ o) THG in the YZ plane. The input face of the crystal is coated with DBAR coating for 1064 nm and 532 nm. The output face of the crystal has an anti-reflection coating for 355 nm. Note that although the crystals were cut for harmonic generation of Nd:YAG laser, i.e. 1064 nm fundamental wavelength, they were usable for the Yb:YAG based laser system with fundamental wavelength of 1030 nm as well. Both the crystals are mounted in temperature-controlled ovens designed by the Faculty of Nuclear Sciences and Physical Engineering at the Czech Technical University in Prague, Czechia. These ovens have a temperature tuning range of 20\mbox{-}55°C with long-term stability better than $\pm$0.01°C.

\section{Experimental results and discussion}

\subsection{SHG experiments}

\begin{figure}[h!]
	\centering
	\includegraphics[scale=0.036]{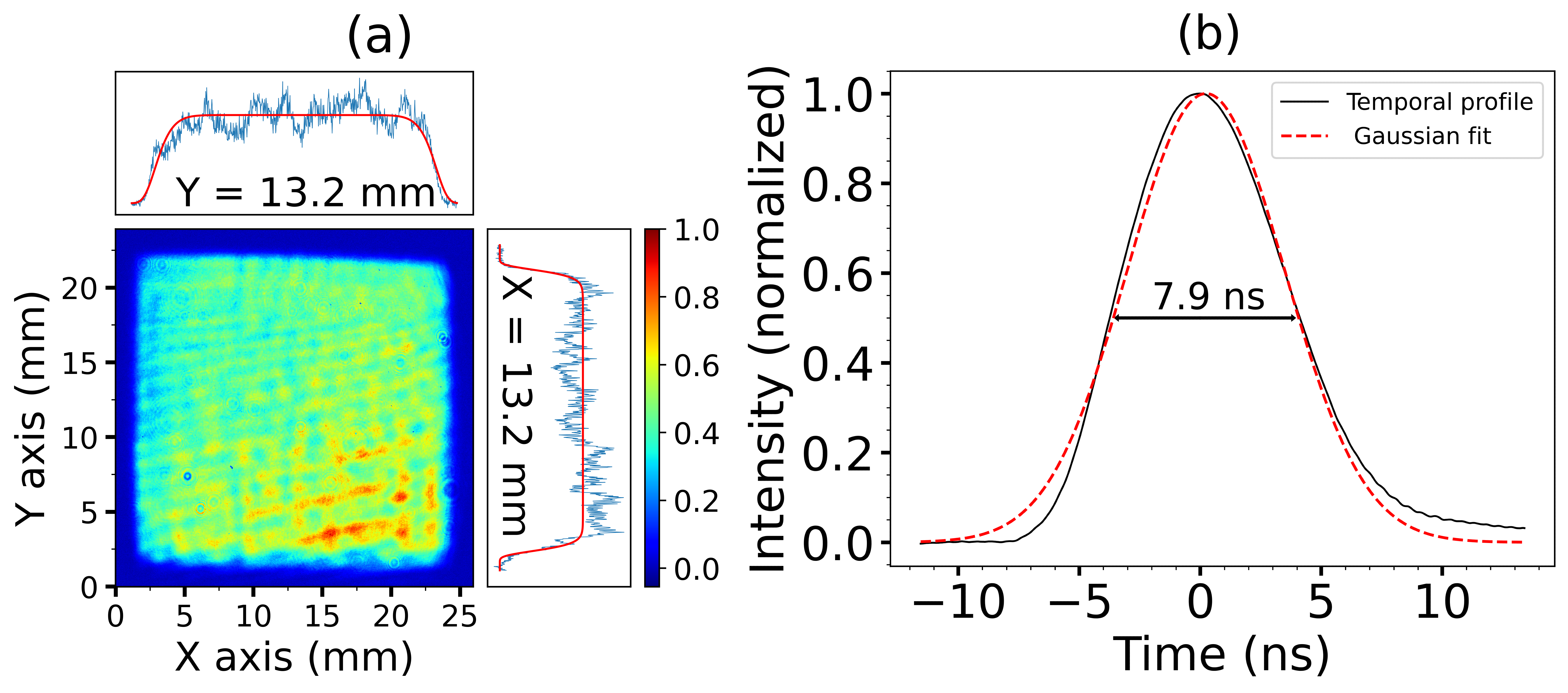}
	\caption{(a) Near-field profile of the input fundamental beam at 7 J. (b) Temporal profile of the fundamental beam.}
	\label{fig:1H_near_field}
\end{figure}

The first set of experiments was focused on SHG. The laser operated at a repetition rate of 10 Hz, delivering pulses with a width of 8 ns and a Gaussian temporal profile. The energy incident on the crystal was slightly over 7 J. Figure~\ref{fig:1H_near_field}(a) shows the spatial profile of energy fluence of the input beam at energy of 7 J. The cross-section near the beam center,
along with its fitted 10th-order super-Gaussian curve, are shown adjacent to it.Note that the beam from the laser is imaged onto the crystal and the near-field camera is in such position that it acquires image corresponding to the crystal position. The beam size on the crystal is roughly 22.1 mm $\times$ 20.2 mm at Full Width at Half Maximum (FWHM). Figure~\ref{fig:1H_near_field}(b) presents the temporal profile, with a measured width of 7.9 ns at FWHM. The LBO crystal was maintained at 35°C, and the pulse energy directed to the crystal was gradually increased up to 7.1 J using the attenuator during the experiment.

\begin{figure}[h!]
	\centering
	\includegraphics[scale=0.35]{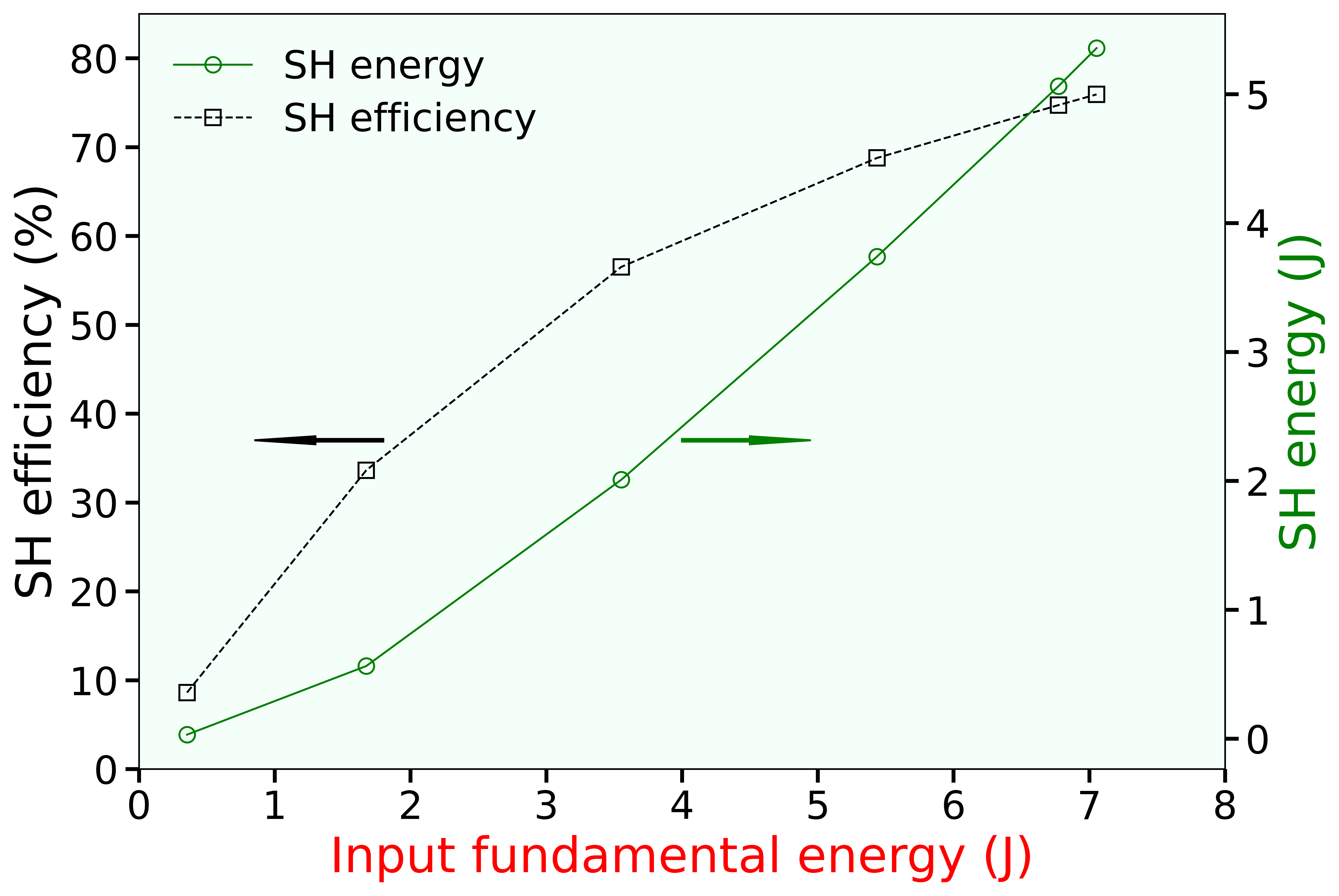}
	\caption{Dependence of the SH pulse energy on the input fundamental pulse energy, and the SHG conversion efficiency.}
	\label{fig:2H_energy_dependence}
\end{figure}

Figure~\ref{fig:2H_energy_dependence} presents the SH pulse energy dependence and the SH conversion efficiency. As the fundamental energy increased, the SH energy rose, reaching the maximum of 5.3 J for the fundamental energy of 7.1 J, corresponding to
a conversion efficiency of 75\%. This represents a record conversion efficiency for a 10-J-class laser system utilizing Gaussian temporal profiles, surpassing the previously reported maximum SHG efficiency of 71.5\% \cite{Sekine:13}.

\begin{figure}[h!]
	\centering
	\includegraphics[scale=0.036]{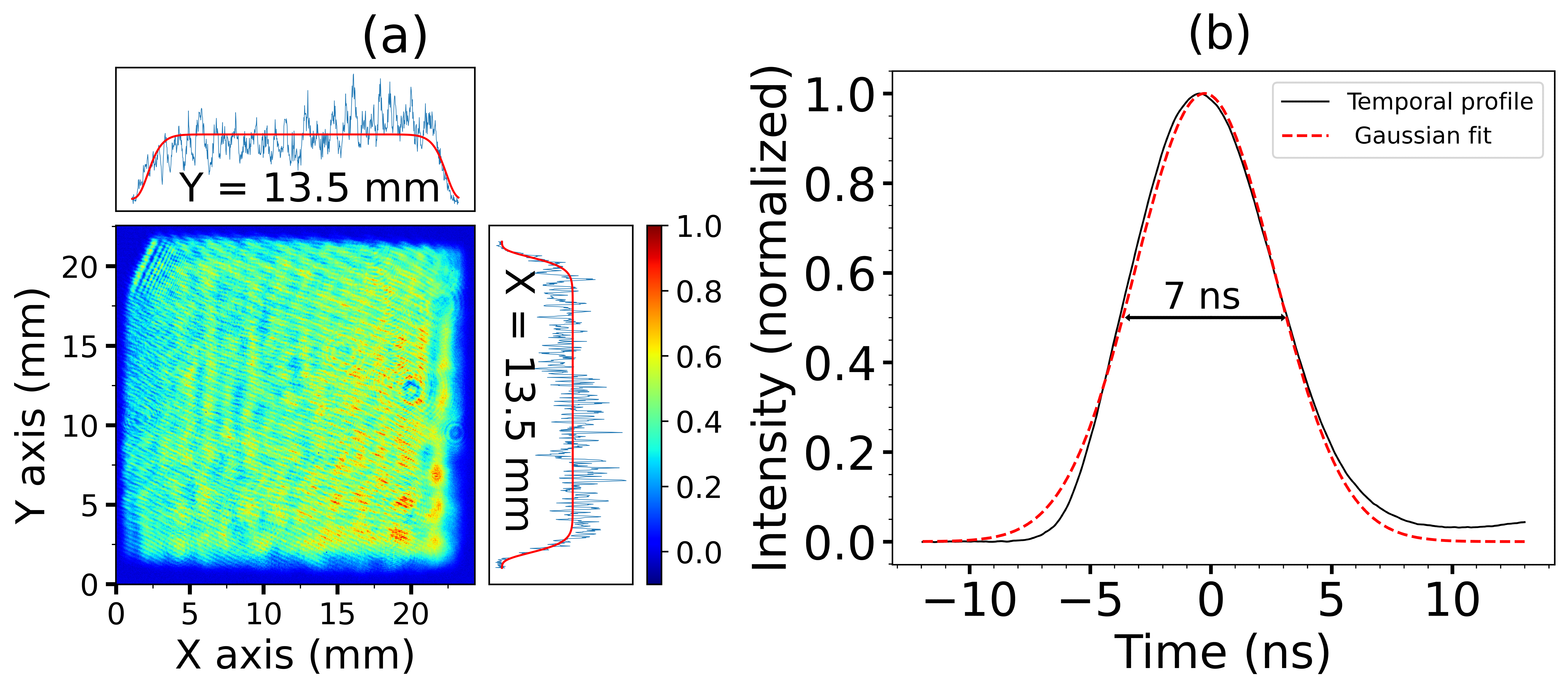}
	\caption{(a) Near-field profile of the SH beam at 5 J. (b) Temporal profile of the SH beam.} 
	\label{fig:2H_near_field}
\end{figure}

\begin{figure}[h!]
	\centering
	\includegraphics[scale=0.036]{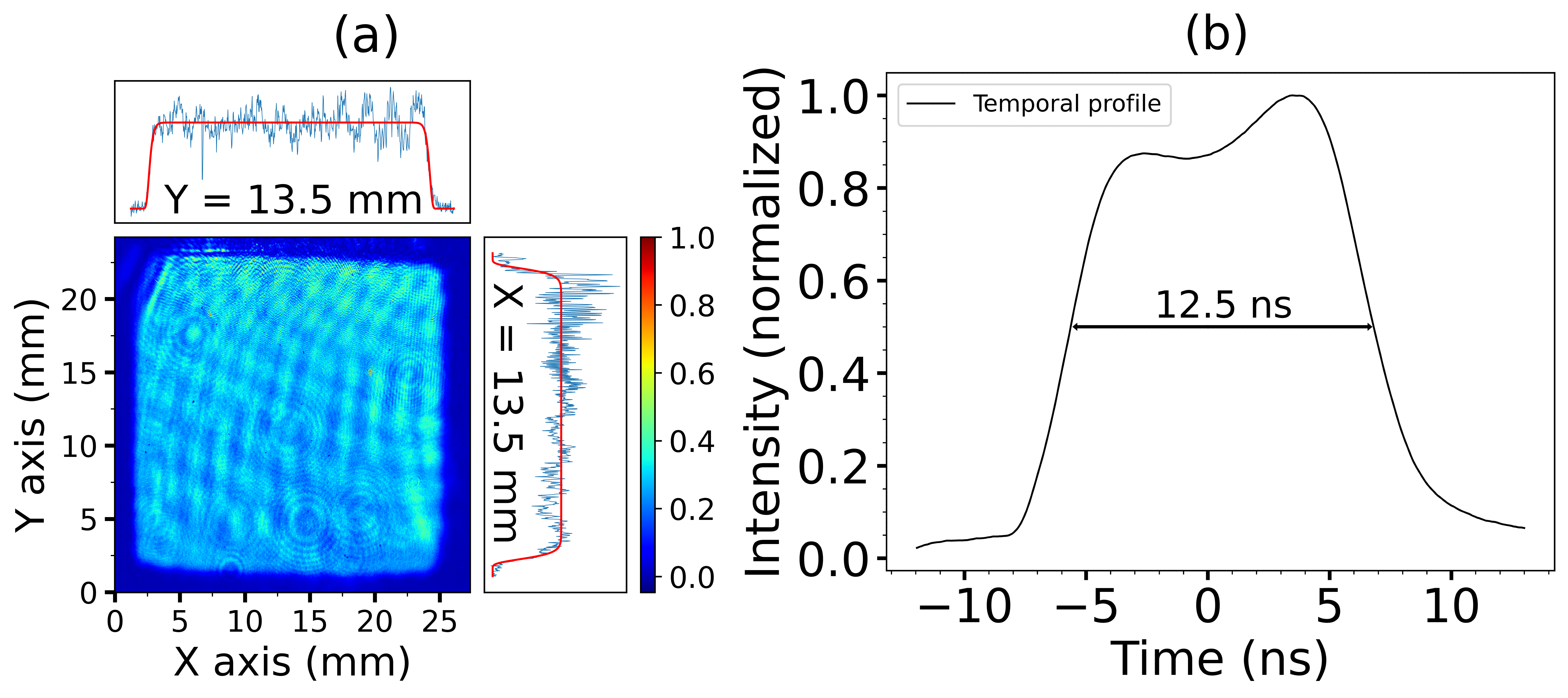}
	\caption{(a) Near-field profile of the unconverted fundamental beam. (b) Temporal profile of the unconverted fundamental beam.}
	\label{fig:1H_near_field_unconverted}
\end{figure}

Figure~\ref{fig:2H_near_field}(a) and Fig.~\ref{fig:1H_near_field_unconverted}(a) depict the near-field energy fluence spatial profiles of the SH and unconverted fundamental beams, respectively. Similarly, Fig.~\ref{fig:2H_near_field}(b) and Fig.~\ref{fig:1H_near_field_unconverted}(b) illustrate the temporal profiles of the SH and unconverted fundamental pulses, respectively. The measured pulse width of the SH is 7 ns (FWHM), i.e. about 1 ns shorter than the fundamental due to the lower conversion efficiency at the lower intensities at the edges of the pulse. The unconverted pulse at the fundamental wavelength has a width of 12.5 \nolinebreak ns, broadened due to pulse depletion around its peak. Note that the asymmetric profile is an artifact of the photodiode’s temporal response. Comparing Fig.~\ref{fig:1H_near_field}(a) and \ref{fig:2H_near_field}(a) reveals the similarity in the near-field profile, indicating that higher intensity areas in the fundamental profile correspond to higher intensity areas in the SH profile.

\begin{figure}[h!]
	\centering
	\includegraphics[scale=0.34]{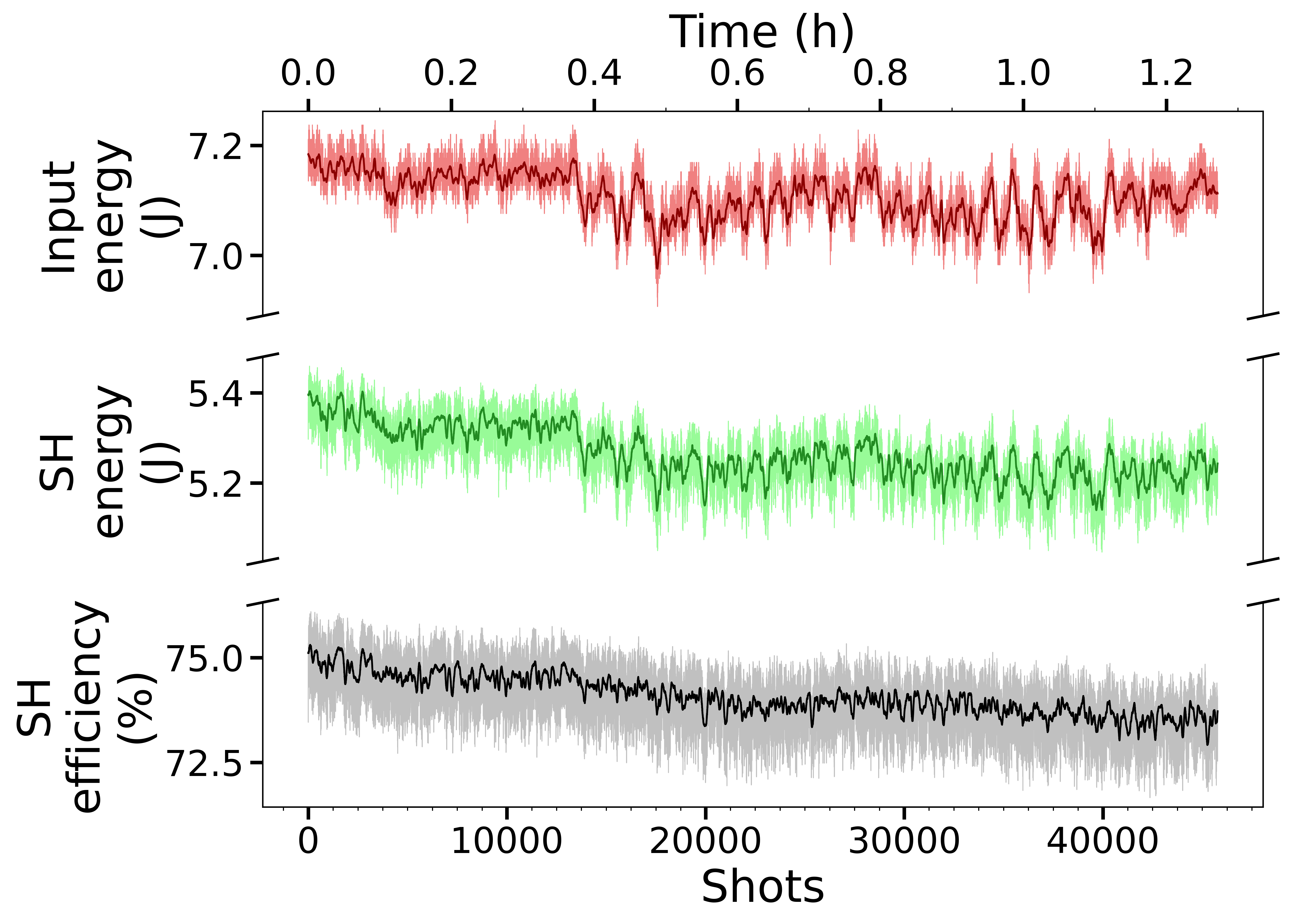}
	\caption{Long-term energy stability of the SH pulse energy.  Light colored narrow lines show pulse to pulse energies whereas the darker thick lines show floating average over 100 pulses.}	
	\label{fig:2H_stability}
\end{figure}

We recorded the long-term pulse energy stability of the SH with its pulse energy exceeding 5 J for a little over 1 hour and 15 minutes, as displayed in Fig.~\ref{fig:2H_stability}. To smooth out fluctuations and highlight longer trends, a floating average of 100 pulses is also shown in the figure. The measured SHG output energy stability was 1.3\% RMS (pulse-to-pulse), compared to 0.7\% RMS (pulse-to-pulse) for the fundamental measured over the same period. The stability of the conversion efficiency was measured at  as 0.96\% RMS (pulse-to-pulse). While the trend of the SHG output pulse energy appears to follow that of the fundamental, the conversion efficiency does not follow the same trend. This discrepancy arises because the SH efficiency starts to be saturated at the given fundamental energy, as indicated in the energy dependence graph in Fig.~\ref{fig:2H_energy_dependence}. Consequently, changes in the fundamental and SH pulse energies do not significantly affect the conversion efficiency. However, a decrease in SH energy and conversion efficiency, along with a decrease in fundamental energy, is observed over time in the figure. The small long term decrease by 1\% of the fundamental pulse energy results in less than 2\% long term decrease of the SH pulse energy.

\begin{figure}[h!]
	\centering
	\includegraphics[scale=0.37]{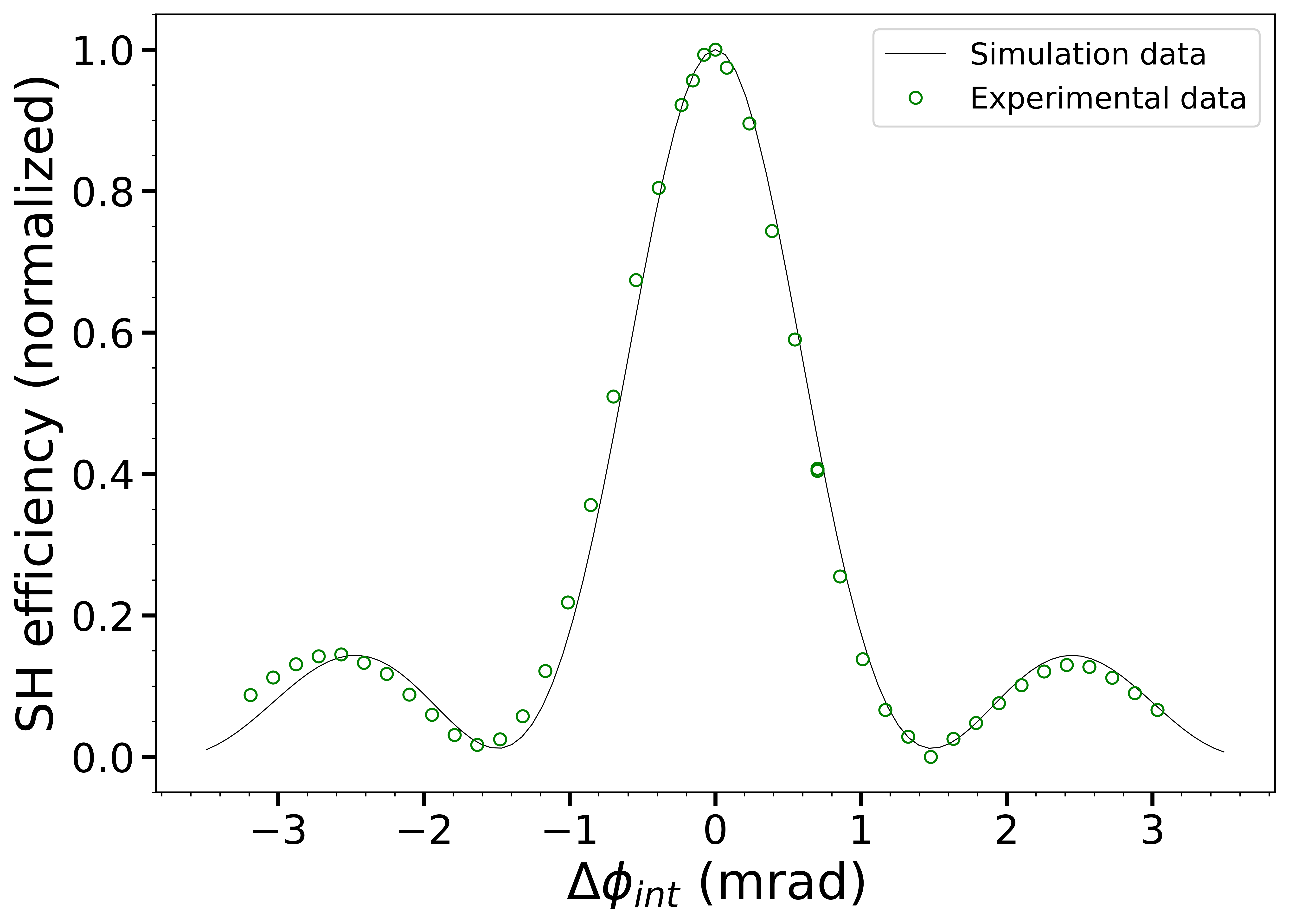}
	\caption{Dependence of the SH conversion efficiency on internal angular detuning for 20 mm long LBO crystal.}	\label{fig:2H_angular_dependene}
\end{figure}

We measured the dependence of the SH pulse energy on the rotation of the LBO crystal about vertical axis. The fundamental energy entering the crystal was set to 6.7 J, and the SH energy was measured while varying the angle ($\phi$), i.e. the angle between the wave vector of the fundamental beam, being in the XY plane of the LBO crystal, and the crystal’s X axis. To support the experimental results, simulations were performed using the mlSNLO software from AS-Photonics~\cite{SNLO}, specifically the 2D-mix-LP module i.e. the module which considers spatial profiles of the beam and long (ns) pulses. The inputs for simulations stemmed from the experiment. The experimental angular changes are related to angles outside the crystal. For comparison with calculated data the experimental angular changes were recalculated into internal angular changes. The refractive index of LBO crystal for the fundamental wavelength in XY plane was calculated using the Sellmeier equation. Subsequently, the internal angles were determined by applying the Snell's law. As shown in Fig.~\ref{fig:2H_angular_dependene}, the internal angular acceptance bandwidths of the SHG in the LBO crystal were approximately 1.33 mrad ($\sim$2.66 mrad·cm) and 1.31 mrad ($\sim$2.62 mrad·cm) at FWHM for the experimental and simulation data, respectively. Hence, the experimental and simulated angular acceptance bandwidths are in good agreement. Also, the width of the calculated curve  at 99\% efficiency is 0.11 mrad, which approximately corresponds to an angular rotation of the crystal by 0.18 mrad. This implies that the external beam pointing fluctuation should be below $\pm$ 0.09 mrad to ensure that the SH conversion efficiency does not decrease by more than 1\%.

\begin{figure}[h!]
	\centering
	\includegraphics[scale=0.37]{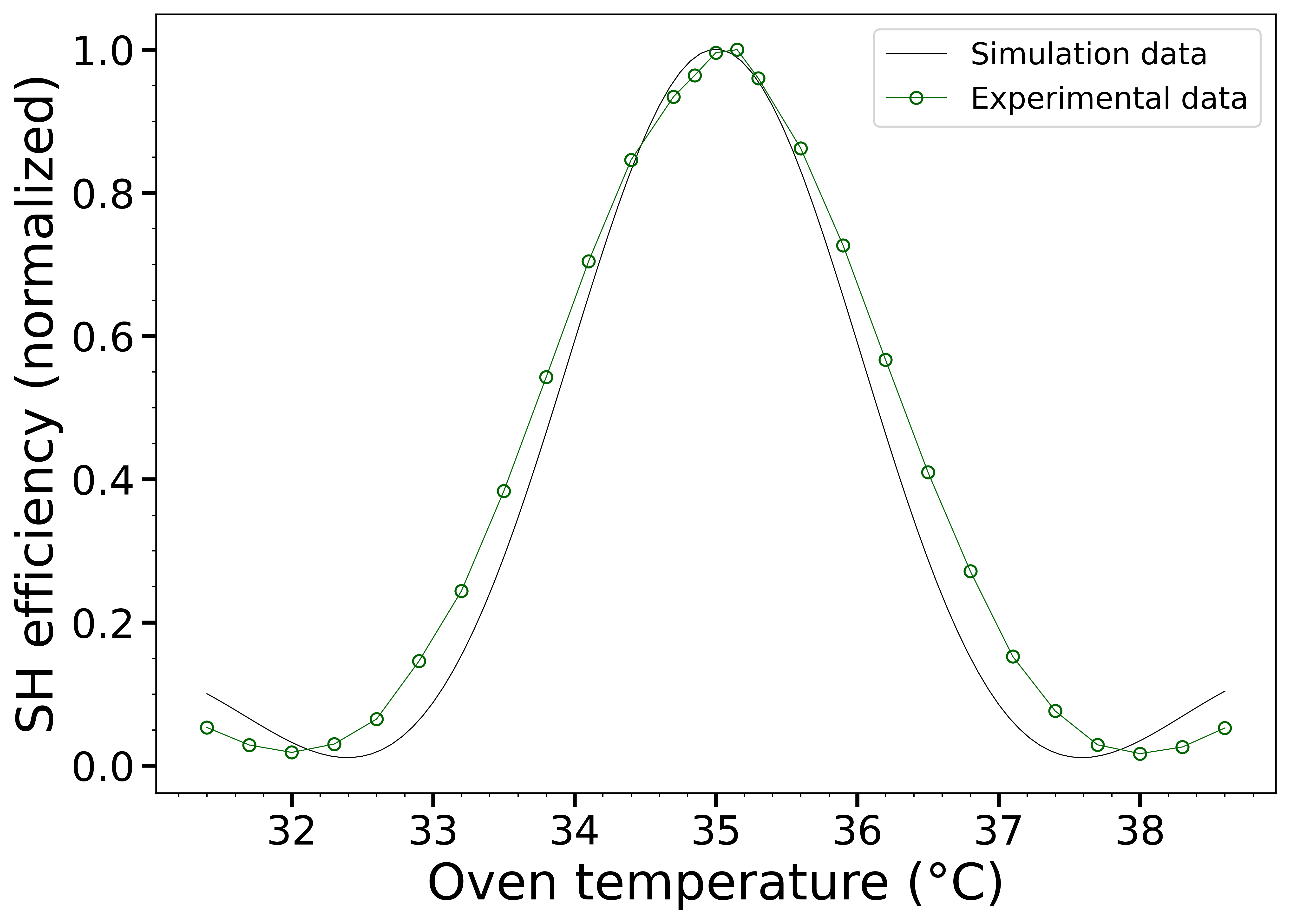}
	\caption{Dependence of the SH conversion efficiency on the oven temperature for the 20 mm long LBO crystal.}	
	\label{fig:2H_temperature_dependene}
\end{figure}

To evaluate the sensitivity of SHG to temperature variations in the LBO crystal, we measured dependence of the SH pulse energy on the LBO crystal temperature. With the fundamental energy maintained at 6.5 J, we recorded the SH pulse energy at various oven temperatures. After each oven temperature change there was enough time for crystal thermalization. To complement our experimental findings, we performed simulations using the 2D-mix-LP module of mlSNLO software for a more comprehensive analysis. As determined from Fig.~\ref{fig:2H_temperature_dependene}, the FWHM temperature acceptance bandwidths for the SHG in the LBO crystal were approximately 2.61 K ($\sim$5.22 K·cm) based on the experimental data and 2.3 K ($\sim$4.6 K·cm) according to the simulation data. The experimental and simulated results  remain in good agreement. The calculated temperature tolerance curve exhibits a width of 0.22 K at 99\% efficiency. Therefore, the oven used for stabilizing the temperature of the LBO crystal during SHG should keep the crystal’s temperature changes within $\pm$ 0.11 K to maintain stable conversion efficiency.

\subsection{THG experiments}

\begin{figure}[h!]
	\centering
	\includegraphics[scale=0.3]{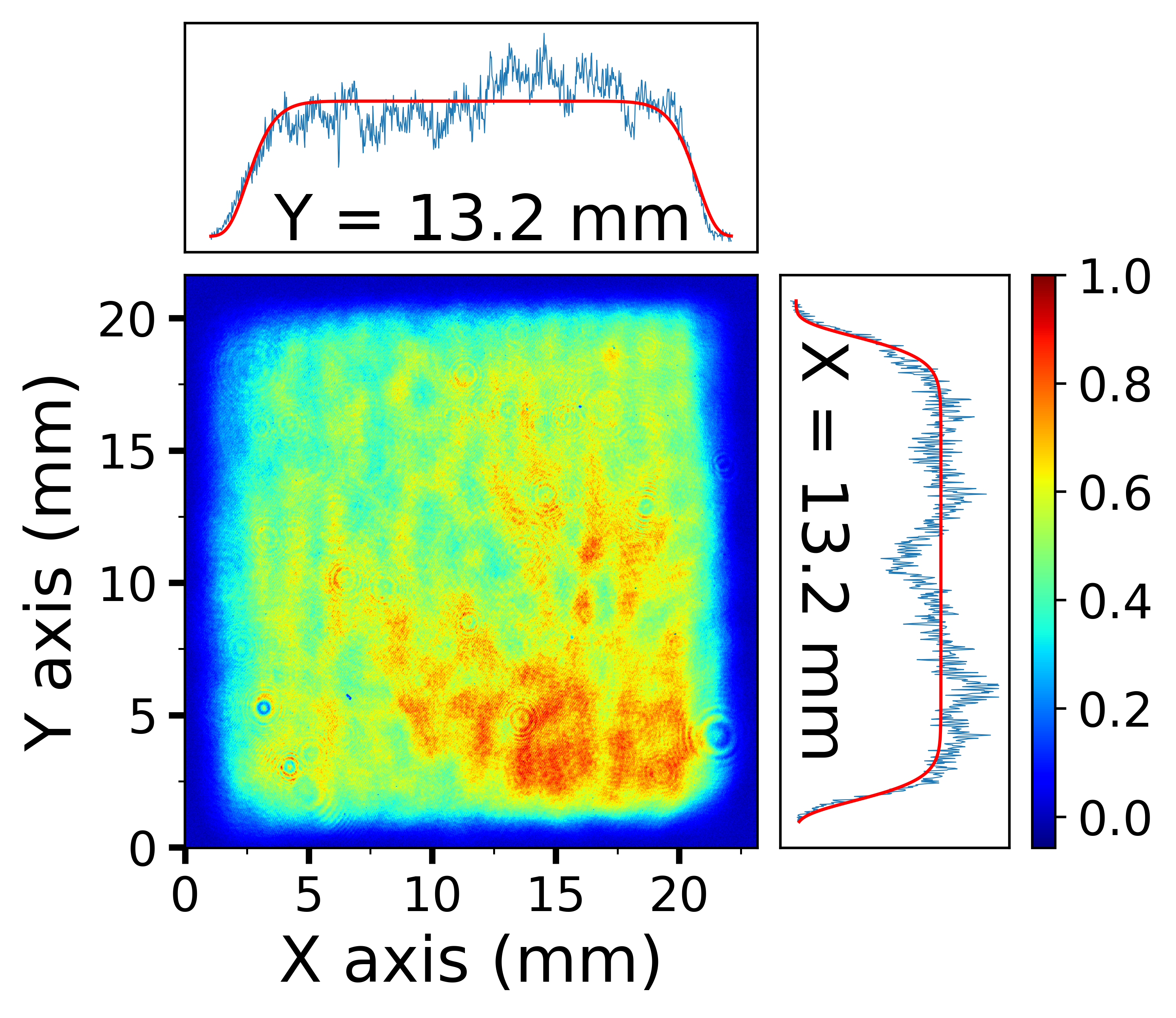}
	\caption{Near-field profile of the input fundamental beam at 6.1 J.} 
	\label{fig:1H_near_field_for_3H}
\end{figure}

\begin{figure}[h!]
	\centering
	\includegraphics[scale=0.35]{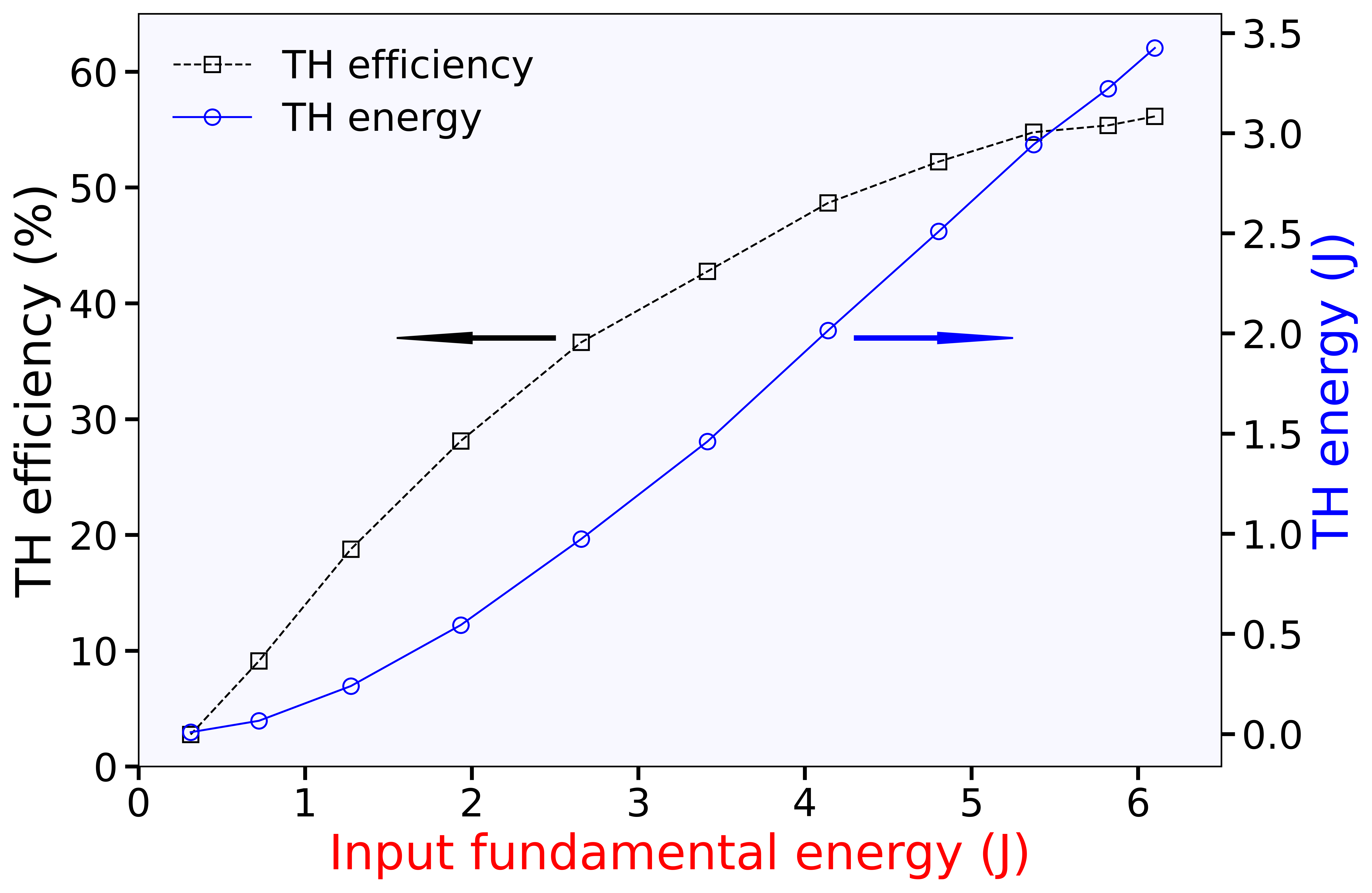}
	\caption{Dependence of the TH pulse energy on the input fundamental pulse energy, and the THG conversion efficiency.}
	\label{fig:3H_energy_dependence}
\end{figure}

\begin{figure}[h!]
	\centering
	\includegraphics[scale=0.036]{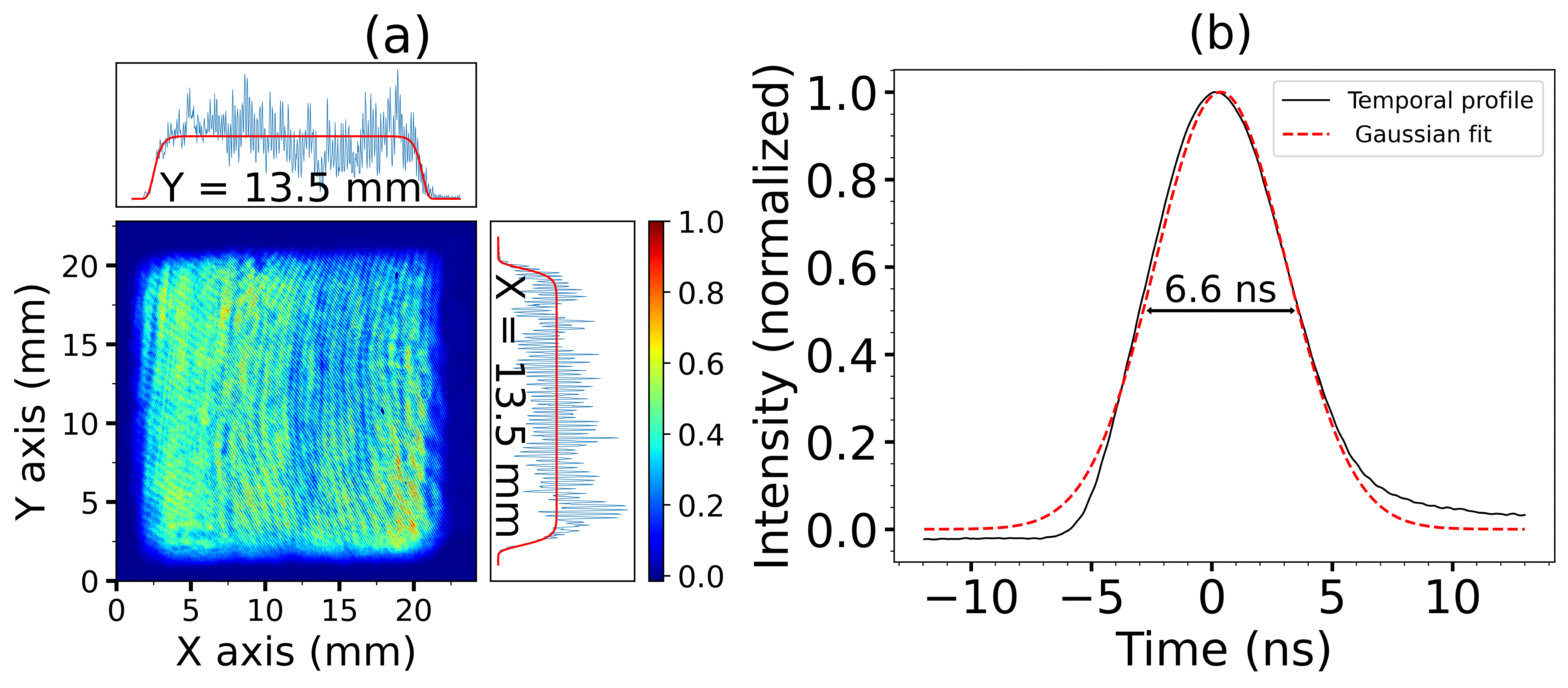}
	\caption{(a) Near-field profile of the TH beam at 3.5 J. (b) Temporal profile of the TH beam.}
	\label{fig:3H_near_field}
\end{figure}

The next set of experiments focused on THG, achieved through sum frequency generation (SFG) between the SH and the unconverted fundamental beams. For these experiments, the fundamental energy was set to 6.1 J, which is lower than the energy used for SHG. This reduction corresponds to the smaller beam dimensions of 19.4 mm $\times$ 18.8 mm compared to those in the SHG experiments, and ensures operation at fluence levels matching those used in the SHG experiments. The beam size reduction utilizing a spatial light modulator \cite{Paliesek_Navrátil_Pilař_Divoký_Smrž_Mocek_2023} was done because the crystal used for THG was optimized for input fundamental wavelengths of 1064 nm, whereas the experiment utilized 1030 nm. Consequently, the crystal was rotated by approximately 12° about the vertical axis to change the $\theta$ angle, i.e. the angle between the crystal’s Z-axis and the wave vector of the input beams in the YZ plane, which reduced its effective aperture,  making beam reduction necessary. The spatial profile of the output at 6.1 J, and its cross-sections near the beam centers fitted with a 10th-order super-Gaussian profile, is shown in Fig.~\ref{fig:1H_near_field_for_3H}. The oven temperature for both SHG and THG crystal was maintained at 35°C. After optimizing the angles  of both the THG and SHG crystals, we achieved an output of 3.5 J at 343 nm with an efficiency of 56\%. The SHG crystal required optimization to ensure that the conversion efficiency was not excessively high, as this would deplete the fundamental beam energy available for subsequent THG. Additionally, the tilt of the THG crystal was carefully optimized to maximize the conversion efficiency for the SFG process. The dependence of THG energy and THG conversion efficiency are illustrated in Fig.~\ref{fig:3H_energy_dependence}. Optimization of both the SHG and THG crystals was performed at the maximum available fundamental energy to ensure the best possible phase-matching conditions. Once optimized, the crystal tilts were kept constant during the subsequent energy dependence measurements to maintain consistency. To maximize conversion efficiency, the mixing ratio of SH to the fundamental pulse energy was set to about 2:1.5, even though the theoretical optimum photon ratio is 2:1. This adjustment was necessary, because after the SHG process, the depleted fundamental output  pulse is longer \cite{1072294}, and its temporal profile transitioned from a Gaussian to a double-Gaussian distribution as shown in Fig.~\ref{fig:1H_near_field_unconverted}(b). As a result, the temporal correlation between the fundamental and SH pulses is not ideal. To compensate for this mismatch, an imbalance in the mixing ratio was introduced. Figure~\ref{fig:3H_near_field}(a) shows the near-field spatial distribution of the TH beam, while Fig.~\ref{fig:3H_near_field}(b) depicts the temporal profile of the TH pulse. The measured pulse width of the TH is 6.6 ns (FWHM).

\begin{figure}[h!]
	\centering
	\includegraphics[scale=0.34]{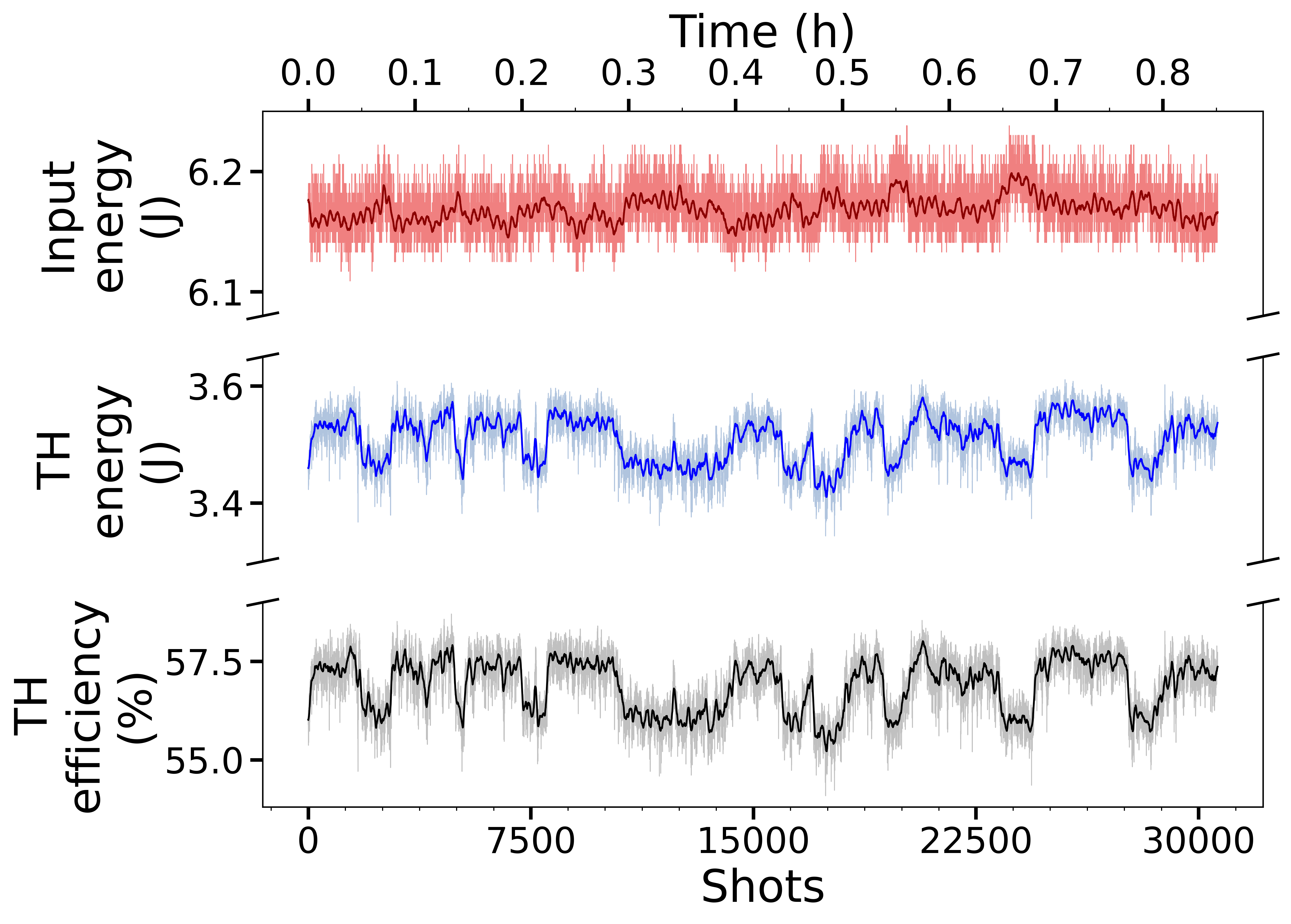}
	\caption{Long-term stability of the TH pulse energy. Light colored narrow lines show pulse to pulse energies whereas the darker thick lines show floating average over 100 pulses.}
	\label{fig:3H_long_term_stability}
\end{figure}

We monitored the THG output energy over more than 30,000 shots (spanning 50 minutes) at a fundamental energy of 6.1 J, as depicted in Fig.~\ref{fig:3H_long_term_stability}. To emphasize longer-term trends, a floating average of 100 pulses is also presented in the figure. The measured stability of the THG output energy was 1.24\% RMS (pulse-to-pulse), compared to 0.26\% RMS (pulse-to-pulse) for the fundamental energy measured over the same period. The stability of the conversion efficiency was measured as 1.29\% RMS (pulse-to-pulse). Both the TH pulse energy and efficiency trends appear to follow a similar pattern, although they do not seem to align with the trend of the input fundamental energy. However, upon closer examination, an inverse relationship can be partially observed. This discrepancy may be partly due to an increase in SHG as the fundamental energy rises, leading to an imbalance in the mixing ratio of SH and unconverted fundamental energy for THG. The excess of SH and the reduction in the fundamental component lead to the decrease in TH output.

\begin{figure}[h!]
	\centering
	\includegraphics[scale=0.37]{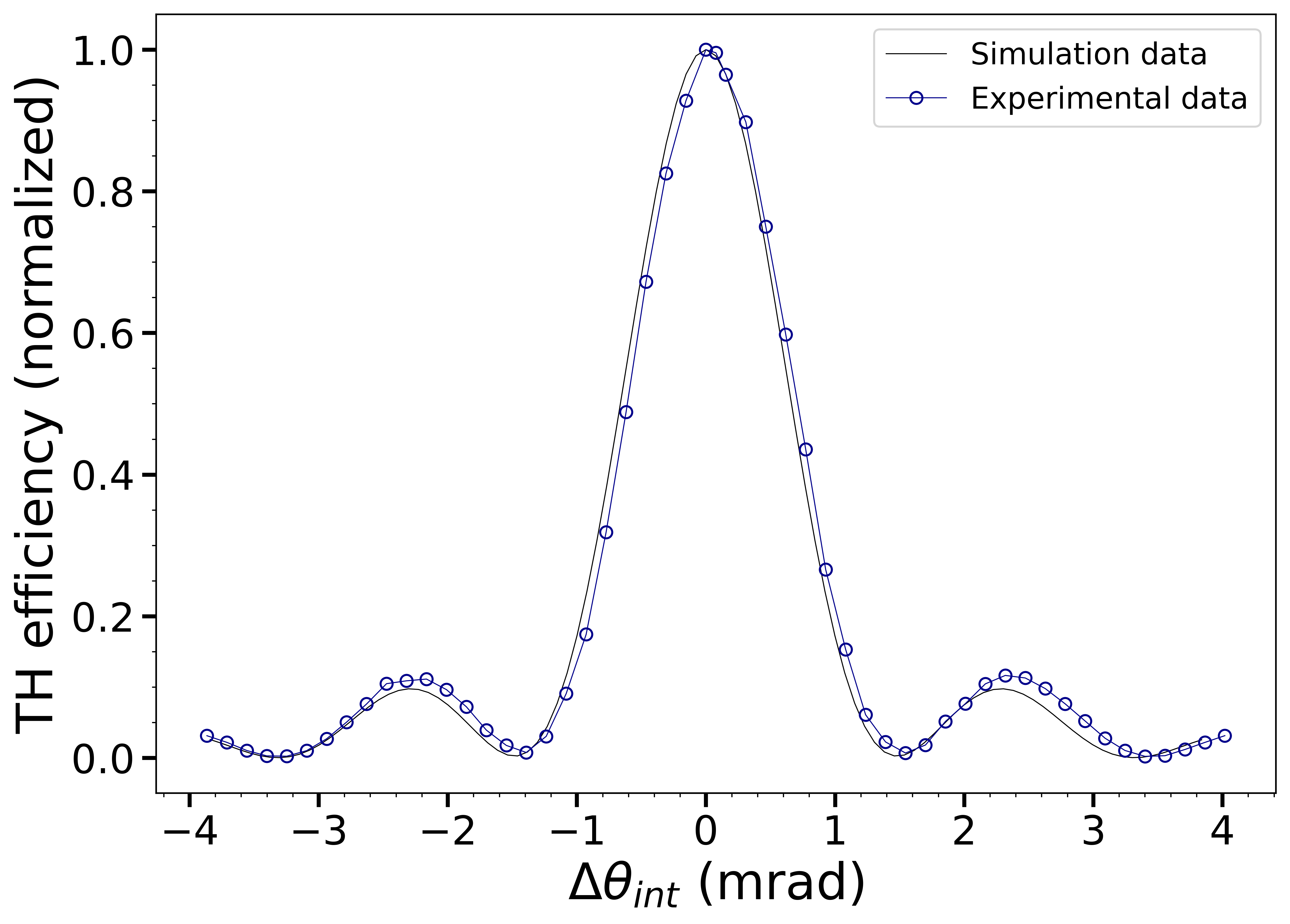}
	\caption{Dependence of the TH conversion efficiency on internal angular detuning for 20 mm long LBO crystal.}	
	\label{fig:3H_angular_dependene}
\end{figure}

The angular acceptance of the THG was determined in a similar way to SHG by measuring the output pulse energy as a function of the phase matching angle of the THG crystal. The fundamental pulse energy input into the crystal was set to 4.7 J, and the TH pulse energy was measured while varying the angle ($\theta$). The results are presented in Fig.~\ref{fig:3H_angular_dependene}, alongside the simulation results obtained using the 2D-mix-LP module of the mlSNLO software. Note that the angular changes are considered inside the crystal. The angular acceptance bandwidths for the THG in the LBO crystal were found to be 1.19 \nolinebreak mrad ($\sim$2.38 mrad·cm) at FWHM for the experimental data and 1.29 mrad ($\sim$2.58 mrad·cm) for the simulation data, demonstrating a good agreement between the two. Additionally, the width of the calculated curve at 99\% efficiency is 0.07 mrad, which roughly translates to an angular crystal rotation of 0.11 mrad. This suggests that to maintain the TH conversion efficiency within a 1\% variation, the external beam pointing fluctuation must remain below $\pm$ \nolinebreak 0.055 mrad.

\begin{figure}[h!]
	\centering
	\includegraphics[scale=0.37]{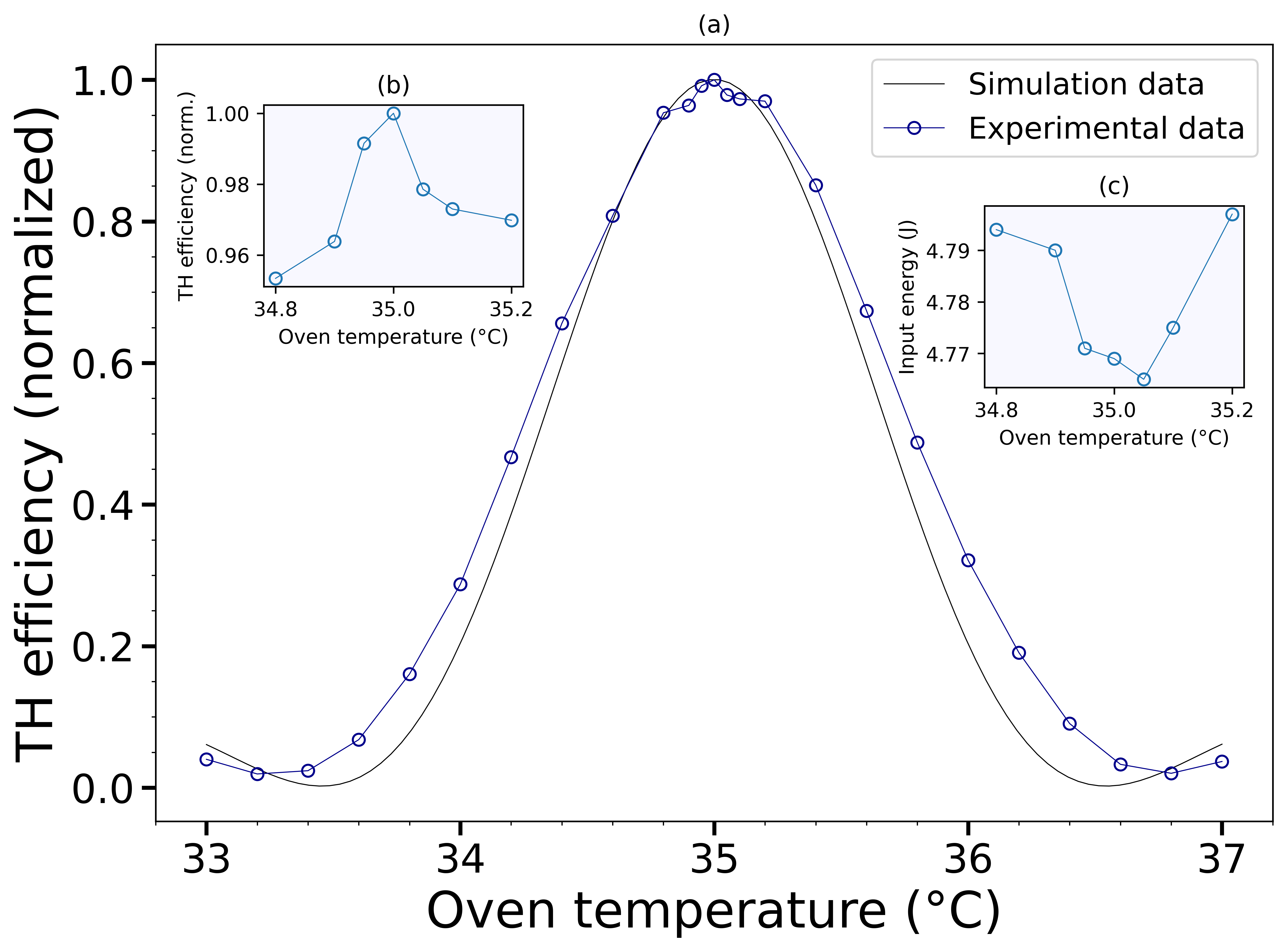 }
	\caption{(a) Dependence of the TH conversion efficiency on the oven temperature for the 20 mm long LBO crystal. (b) Dependence of the TH conversion efficiency on the oven temperature with a reduced range of temperature points on the x-axis. (c) Input fundamental energy measured at the different oven temperature as it enters the crystal.}	
	\label{fig:3H_temperature_dependene}
\end{figure}

Lastly, the THG process at 343 nm was investigated over a temperature range of 33-37 °C to determine the temperature acceptance bandwidth. With the fundamental energy kept constant at 4.7 J, we measured the resulting TH pulse energy. Additionally, simulations were conducted using the 2D-mix-LP module of mlSNLO to provide a more detailed analysis. As shown in Fig.~\ref{fig:3H_temperature_dependene}(a), the temperature acceptance bandwidths for THG in the LBO crystal were approximately 1.35 K ($\sim$2.70 K·cm) based on experimental data and 1.55 K ($\sim$3.1 K·cm) according to simulation results, both measured at FWHM. The experimental and simulated results are in good agreement. Also, the inconsistency observed at the peak of the curve can be explained by examining Fig.~\ref{fig:3H_temperature_dependene}(b), Fig.~\ref{fig:3H_temperature_dependene}(c). A comparison between Fig.~\ref{fig:3H_temperature_dependene}(b) and Fig.~\ref{fig:3H_temperature_dependene}(c) reveals that as the input fundamental energy decreases, the conversion efficiency increases. However, this trend is not strictly proportional, likely due to fluctuations in beam pointing, and other instabilities which ultimately causes the peak of the curve to appear irregular. Furthermore, the calculated temperature tolerance curve has a width of 0.19 K at 99\% efficiency. Thus, to ensure optimal conversion efficiency during THG, the oven used for stabilizing the LBO crystal's temperature must maintain its temperature within $\pm$ \nolinebreak 0.095 \nolinebreak K or better.

\section{Conclusion}

In this study, we explored the processes of SHG and THG using a LBO crystals under controlled experimental conditions. For SHG, we achieved a peak conversion efficiency of 75\%, yielding an output of 5.3 J at 515 nm. The output energy of SHG demonstrated a stability of 1.3\% RMS, with a conversion efficiency stability of 0.96\% RMS, for the fundamental energy stability of 0.7\% RMS. The measured angular and temperature acceptance bandwidths for SHG closely matched the simulations.

In the case of THG, we obtained an output of 3.5 J at 343 \nolinebreak nm, with a conversion efficiency of 56\%. The THG process was notably sensitive to the mixing ratio of the SH and residual fundamental pulses. The output energy stability for THG was measured at 1.24\% RMS, with a conversion efficiency stability of 1.29\% RMS, for the fundamental energy stability of 0.26\% RMS. The angular and temperature acceptance bandwidths for THG were consistent with the simulated predictions although there is a small discrepancy in temperature acceptance bandwidth.

Overall, the results confirm the efficacy of LBO crystals for SHG and THG for high-energy high-power lasers, with most findings aligning well with theoretical models. Furthermore, the conversion efficiency could be increased by using a super-Gaussian temporal profile. Additionally, the angular acceptance bandwidths for SHG and THG at FWHM exceed 2 mrad·cm, while the temperature acceptance bandwidth at FWHM surpasses 2.5 \nolinebreak K·cm, both of which are notably large. We also demonstrate the long-term stability of these processes in LBO crystals, affirming the robustness of the frequency conversion. These findings collectively highlight LBO as an excellent crystal choice for efficient harmonic generation in high-energy, high-average-power laser applications. An important advantage demonstrated in this work is the ability to automatically switch between SHG and THG outputs without requiring physical realignment of the setup. This flexibility greatly enhances operational convenience and makes the system particularly attractive for applications requiring rapid switching between harmonics. Additionally, we have developed a versatile system that can be easily adapted with only minor adjustments to accommodate both Nd:YAG and Yb:YAG lasers, thereby broadening its applicability across diverse laser platforms.

\begin{center}\textbf{Acknowledgement}\end{center}

\noindent This work was supported by Technology Agency of the Czech Republic (TM03000045). This work was co-funded by the European Union and the state budget of the Czech Republic under the project LasApp  (CZ.02.01.01/00/22\textunderscore008/0- \\ 004573).

The authors declare no conflicts of interest.

\bibliography{hpl-template}

\end{document}